\documentstyle[11pt,amstex,epsfig]{article}

\newcommand{\dd}{\text{d}}

\newcommand{\ee}{\text{e}}

\newcommand{\p}{\partial}

\newcounter{exerc}
\begin{document}

\title{A universality class\\
in Markovian persistence}

\author{O. Deloubri{\`e}re \\
Laboratoire de Physique Th{\'e}orique\footnote{Laboratoire associ{\'e} au Centre National de la Recherche Scientifique - UMR 8627}\\
B{\^a}timent 210\\
Universit{\'e} de Paris-Sud\\
91405 Orsay cedex, France\\}

\date{\today}

\maketitle

\begin{abstract}
\noindent 
We consider the class of Markovian processes defined by the equation $\dd x /\dd t = -\beta x + \sum_k z_k \delta (t-t_k)$. Such processes are encountered in systems (like coalescing systems) where dynamics creates discrete upward jumps at random instants $t_k$ and of random height $z_k$. We observe that the probability for these processes to remain above their mean value during an interval of time $T$ decays as $\ee ^{-\theta T}$ defining $\theta$ as the persistence exponent. We show that $\theta$ takes the value $\beta$ which thereby extends the well known result of the Gaussian noise case to a much larger class of non-Gaussian processes.

\vspace{.2cm}
\noindent {\bf PACS number(s): 05.40+j}
\end{abstract}

\vspace{.2cm}
\noindent L.P.T. -- ORSAY 00/40

\newpage

\section{Introduction}

Let us consider a stationary random process $x(t)$ and suppose that at $t=0$ this process is above a given threshold $X$. What is the probability $Q(T)$ that the process has never crossed the threshold $X$ up to time $t=T$ ? In many examples of statistical mechanics we observe
\begin{equation}
Q(T) \sim \ee ^{-\theta_{X} T} ,\qquad T \to \infty
\end{equation}
where $\theta_X$ is a function of the threshold. The question of determining $\theta_X$ is called the "persistence problem" and has recently been the main interest of many papers in the study of out of equilibrium systems. For a short review the reader can see Ref.\,\cite{majum} and references therein. In many persistence problems, only the case where $X=\langle x \rangle$ is considered and usually one writes $\theta$ for $\theta_{\langle x \rangle}$.

It often happens that stationary processes can be obtained from physical nonstationary systems by rescaling the space variable and taking the new time variable $t=\log \tau$ \cite{majum}, where $\tau$ is the physical time. In terms of the latter the persistence probability decays as a power law, thus defining $\theta_X$ as a new independant exponent \cite{MBCS} called the persistence exponent.

In this paper we present a class of stationary Markov processes $x(t)$ that share a universal persistence exponent when $X=\langle x \rangle$. This class of processes is defined by the linear equation of motion
\begin{equation}
\frac{\dd x}{\dd t} = - \beta x + \eta (t)
\label{langevin}
\end{equation}
where $\eta (t)$ is a noise term consisting of upward jumps uniformly distributed along the time axis with density $\rho$,
\begin{equation}
\eta(t) = \sum_{k=-\infty}^{\infty} z_k \delta (t-t_k)
\label{noisedelta}
\end{equation}
The jump heights $z_k$ are independent identically distributed random variables of law $R(z)$ such that $\langle z\rangle$ exists.

If the jump heights are all equal to $a$, the limit $a \to 0$, $\rho \to \infty$ with $\rho a^2$ fixed transforms the shifted noise $\eta(t) - \rho a$ into Gaussian white noise. In this case Eq.\,(\ref{langevin}) is the usual Langevin equation, which is therefore included as a limit in our class of processes.

The interest of this work is to solve the persistence problem for the process defined by Eq.\,(\ref{langevin}). Simulations have shown that for $T$ large $Q(T)$ decays exponentially with an inverse decay rate $\theta_X$ depending on the threshold and on the shape of $R(z)$, such as a Dirac peak, an exponential, or a power law. However if $X=\langle x \rangle = \rho \langle z \rangle / \beta$ the result $\theta=\beta$ appears for any law $R(z)$. This equality was known, until now, only for the Langevin equation, with Gaussian $\eta(t)$ \cite{MBCS,SMR,OCB}.


\section{A simple method for the calculation of $\theta$}
\label{secjumpnoise}

In this section we show that $\theta_X$ is the lowest eigenvalue of a differential operator. Then when the threshold is set to $X = \langle x \rangle$, the persistence problem is solved straightforwardly.

\subsection{Transformation into an eigenvalue problem}
Let $P^*(x,t)$ be the probability distribution of $x$ at time $t$. It is the solution of the master equation
\begin{equation}
\p_{t} P(x,t)=\beta \p_{x} (x P(x,t))
+ \rho \int_{0}^{\infty} P(x-z,t) R(z) \, \dd z - \rho P(x,t)
\label{Mastereqn}
\end{equation}
For the persistence problem only the subclass of functions $x(t)$ that are greater than $X$ during the time interval $T$ is relevant. Let $P^*_{+}(x,t)\dd x$ be the probability that $x(t')$ has not crossed the level $X$ for $t' \in \left[ 0,t\right]$ (given that $x(0) > X$) and that $x(t)$ is in the interval $\left[ x,x + \dd x\right]$. $P^*_{+}(x,t)$ is then solution of the same master equation but with the absorbing boundary condition $P^*_{+}(x,t)=0$ when $x<X$. Because of this condition, $P^*_{+}(x,t)$ will decay to zero as $t \to \infty$. We will write it as a superposition of decay modes $P^{\lambda}_{+}(x) \ee ^{-\lambda_X t}$, where $P^{\lambda}_{+}(x)$ is an eigenfunction satisfying the eigenvalue equation corresponding to Eq.\,(\ref{Mastereqn})
\begin{equation}
-\lambda_X P^{\lambda}_{+}(x)=\beta\frac{\dd}{\dd x} (x P^{\lambda}_{+}(x))
+ \rho \int_{0}^{\infty} P^{\lambda}_{+}(x-z) R(z) \, \dd z - \rho P^{\lambda}_{+}(x)
\label{MastereqnVP}
\end{equation}
with the condition $P^{\lambda}_{+}(x)=0$ when $x<X$. Let $\lambda_X^*$ be the eigenvalue of the slowest decaying mode. Then we have
\begin{equation}
Q(T) \sim \ee ^{-\lambda_X^* T} \int_{X}^{\infty} P^{\lambda^*}_+(x) \, \dd x \qquad \hbox{as} \quad T \to \infty
\end{equation}
so that
\begin{equation}
\theta_X=\lambda^*_X
\label{defpersist}
\end{equation}
if $\int_X^{\infty} P^{\lambda^*}_+(x) \dd x$ exists.

Finally let us call $P^{\theta}_{+}(x) \equiv P^{\lambda^*}_{+}(x)$ the persistence mode. It is necessary to have
\begin{equation}
\begin{split}
P^{\theta}_+(x) \geq 0 \qquad \hbox{for} \quad x \geq X \quad
\hbox{and} \quad \int_{X}^{\infty} P^{\theta}_+(x) \dd x < \infty
\end{split}
\label{physicalsolution1}
\end{equation}
in order to define $P^{\theta}_+ (x,t)$ as a probability distribution. Moreover the definition of the persistence problem ensures that $x(0)$ is chosen within the stationary distribution of the process. This distribution is such that $\langle x \rangle$ exists. Therefore we will only focus on solutions $P^{\theta}_+ (x)$ with
\begin{equation}
\int_{X}^{\infty} \, x \, P^{\theta}_+(x) \dd x < \infty
\label{physicalsolution2}
\end{equation}

As a result solving the persistence problem is now equivalent to calculating the lowest eigenvalue $\lambda^*_X$ of Eq.\,(\ref{MastereqnVP}) that satisfies Eqs.\,(\ref{physicalsolution1}) and (\ref{physicalsolution2}).


\subsection{Calculation of $\theta$}

We will now show that $\lambda^*_{\langle x \rangle}=\beta$ irrespective of the jump size distribution. To this end we define
\begin{equation}
f^{(n)}_{\lambda X} = \int_{X}^{\infty}\,x^n P^{\lambda}_{+}(x) \dd x,\qquad n=0,1,\ldots
\label{moments}
\end{equation}
and we integrate Eq.\,(\ref{MastereqnVP}) on $x$ from $X$ to $\infty$.
As $P^{\theta}_{+}(x)=0$ when $x<X$, we obtain the first relation
\begin{equation}
\lambda_{X}f^{(0)}_{\lambda X}=\beta X P^{\lambda}_{+}(X)
\label{eqntheta1}
\end{equation}
We now multiply Eq.\,(\ref{MastereqnVP}) by $x$ and integrate as before.
After one integration by parts we get
\begin{equation}
\lambda_{X} f^{(1)}_{\lambda X} = \beta X^2 P^{\lambda}_{+}(X) + (\beta + \rho) f^{(1)}_{\lambda X} 
- \rho \int_{X}^{\infty}\dd x \int_{0}^{\infty}\dd z\, x P^{\lambda}_{+}(x-z) R(z)
\label{x}
\end{equation}
Using the absorbing boundary condition, we can write
\begin{equation}
\int_{X}^{\infty}\dd x\int_{0}^{\infty}\dd z\, x P^{\lambda}_{+}(x-z) R(z)
= \int_{X}^{\infty}\dd y P^{\lambda}_{+}(y)\int_{0}^{\infty}\dd z\, (y+z) R(z)
\label{xx}
\end{equation}
and the RHS of this equation simply reduces to $f^{(1)}_{\lambda X} + \langle z \rangle f^{(0)}_{\lambda X}$. Then Eqs.\,(\ref{x}) and (\ref{xx}) give the second relation
\begin{equation}
\lambda_{X} f^{(1)}_{\lambda X} = \beta X^2 P^{\lambda}_{+}(X) + \beta f^{(1)}_{\lambda X} - \rho \langle z \rangle f^{(0)}_{\lambda X}
\label{eqntheta2}
\end{equation}
We then eliminate $P^{\lambda}_{+}(X)$ from Eqs.\,(\ref{eqntheta1}) and (\ref{eqntheta2}) and obtain finally using $\rho \langle z \rangle = \beta \langle x \rangle$
\begin{equation}
\left(f^{(1)}_{\lambda X}-Xf^{(0)}_{\lambda X}\right)\lambda_{X} = \beta \left(f^{(1)}_{\lambda X}- \langle x \rangle f^{(0)}_{\lambda X}\right)
\label{eqntheta}
\end{equation}
which is valid for any $\lambda_X$ for which $f^{(1)}_{\lambda X} < \infty$. One can remark that Eqs.\,(\ref{eqntheta1}) and (\ref{eqntheta2}), and therefore Eq.\,(\ref{eqntheta}), could not be obtained if the jump heights were not all positive. So this property is essential in the present work.

Let us now take $X$ equal to the average of the process: $X=\langle x \rangle$. Eq.\,(\ref{eqntheta}) shows that we must have
\begin{equation}
\lambda_{\langle x \rangle}=\beta \qquad \hbox{or} \qquad f^{(1)}_{\lambda \langle x \rangle} / f^{(0)}_{\lambda \langle x \rangle} = \langle x \rangle 
\label{solution}
\end{equation}
If in Eq.\,(\ref{solution}) we set $\lambda_{\langle x \rangle}=\lambda^*_{\langle x \rangle}=\theta$, we see from Eqs\,(\ref{physicalsolution1}) and (\ref{moments}) that $f^{(1)}_{\theta \langle x \rangle} / f^{(0)}_{\theta \langle x \rangle}$ is the average of the process $x$ in the persistence mode, \emph{i.e.}, given that it remains always above $X=\langle x \rangle$. This quantity is therefore larger than $\langle x \rangle$ and the second one of  Eqs.\,(\ref{solution}) cannot be satisfied. Finally the only possibility is  $\lambda^*_{\langle x \rangle}=\beta$ which proves the main result of this note.

For $\lambda _{\langle x \rangle}\neq \lambda^*_{\langle x \rangle}$ Eq.\,(\ref{solution}) shows that
\begin{equation}
\int_{\langle x \rangle}^{\infty} \dd x \, (x - \langle x \rangle)P^{\lambda}_{+}(x) = 0 \qquad \qquad \lambda_{\langle x \rangle} > \lambda^*_{\langle x \rangle}
\label{eqnonezero}
\end{equation}
It follows that for $\lambda_{\langle x \rangle} \neq \lambda^*_{\langle x \rangle}$ any eigenfunction $P^{\lambda}_{+}(x)$ with $f^{(1)}_{\lambda X} < \infty$ must have at least one zero for $\langle x \rangle < x < \infty$. One can also see Eq.\,(\ref{eqnonezero}) as an orthogonality relation between two eigenvectors of Eq.\,(\ref{MastereqnVP}) that are $x - \langle x \rangle$ (with the eigenvalue $-2 \beta$) and $P^{\lambda}_{+}(x)$.

One final remark is necessary. We have studied the persistence problem in the most natural situation, that is when the condition $f^{(1)}_{\lambda \langle x \rangle} < \infty$ is fulfilled. However one could be interested in other cases where the initial distributions $P(x,0)$ has a large tail for $x \to \infty$ that do not ensure this condition. If one starts with such a distribution, one can expect the persistence problem to be governed by the shape of this initial distribution, with different answers depending on the choice of the latter. Therefore large initial distributions where $f^{(1)}_{\lambda X} = \infty$ are simply cases of no interest in our study.


\subsection{Perturbation for $X$ close to $\langle x \rangle$}

Very few cases in statistical physics offer the possibility of calculating the quantity
\begin{equation}
{\langle x \rangle}_X = f^{(1)}_{\theta X}/f^{(0)}_{\theta X}
\end{equation}
which is the average of the process in the persistence mode.
However as $f^{(n)}_{\theta \langle x \rangle}=f^{(n)}_{\beta \langle x \rangle}$ one can assume that
\begin{equation}
f^{(n)}_{\theta X} = f^{(n)}_{\beta \langle x \rangle} + \tilde{f}^{(n)}_{\theta X}
\end{equation}
where $\tilde{f}^{(n)}_{\theta X} \to 0$ if $|X-\langle x \rangle| \to 0$. Rewriting Eq.\,(\ref{eqntheta}) for $\lambda_X=\lambda^*_X$ one gets
\begin{equation}
\theta_X=\beta \frac{f^{(1)}_{\theta X}- \langle x \rangle f^{(0)}_{\theta X}}{f^{(1)}_{\theta X}-Xf^{(0)}_{\theta X}}
\end{equation}
which can also be written
\begin{equation}
\theta_X=\beta \frac{1}{1-\frac{X-\langle x \rangle}{{\langle x \rangle}_X - \langle x \rangle}}
\label{eqntheta3}
\end{equation}
Finally one can suppose that $X$ is close to $\langle x \rangle$ and perturbs Eq.\,(\ref{eqntheta3}) to find
\begin{equation}
\theta_X=\beta (1 + \frac{X-\langle x \rangle}{{\langle x \rangle}_{\langle x \rangle}-{\langle x \rangle}} + \ldots) \qquad |X-\langle x \rangle| \to 0
\label{thetapert}
\end{equation}
Unfortunately we cannot go further here. We do not know how to calculate ${\langle x \rangle}_{\langle x \rangle}$ except in the Gaussian limit which will be the focus of next section.


\section{The special case of Gaussian noise}
\label{secGaussian}

Our goal here is to illustrate the method of Sec. 2 in the special known case of Gaussian process. If one considers constant jumps with $R(z)=\delta(z-a)$ and sets $\rho a^2=\Gamma$, one can take the limit $a \to 0$ with $\Gamma$ fixed in Eq.\,(\ref{Mastereqn}) and find
\begin{equation}
\p_{t} P(x,t)=\beta \p_{x} ((x-\langle x \rangle) P(x,t))
+ \frac{\Gamma}{2}\p_{xx}P(x,t) + {\cal O}(a)
\label{FokkerPlanck}
\end{equation}
In this limit Eq.\,(\ref{FokkerPlanck}) leads to the Fokker-Planck equation for the zero average process $x(t)-\langle x \rangle$ which hence becomes the stationary Gaussian Markovian process also called the Ornstein-Uhlenbeck process. This shows that the Gaussian case, which is of high interest in statistical physics, is included as a special limit in the class of processes studied in the present work.


\subsection{Calculation of $\theta$}

For simplicity we will now write $x$ instead of $x-\langle x \rangle$ and $X$ instead of $X-\langle x \rangle$. We thus consider the zero average Gaussian process $x(t)$ and we want to determine the persistence exponent $\theta_X$ related to the probability of having never crossed the level $X$ upto time $T$.
As in Sec. \ref{secjumpnoise} it is equivalent to find the lowest eigenvalue $\lambda_X^*$ of 
\begin{equation}
-\lambda_{X} P^{\lambda}_{+}(x) =\beta\frac{\dd}{\dd x} (x P^{\lambda}_{+} (x))
+ \frac{\Gamma}{2}\frac{\dd^2}{\dd x^2}P^{\lambda}_{+} (x) \, , \qquad P^{\lambda}_{+}(X)=0
\label{FokkerPlanckVP}
\end{equation}
The solution of this equation is
\begin{equation}
P^{\lambda}_{+} (x)=\ee ^{-y^2/4}(A_1 D_{\lambda/\beta}(y) + A_2 D_{\lambda/\beta}(-y))
\end{equation}
where $y=\sqrt{2\beta/\Gamma}x$ is dimensionless and $D_{\nu}$ is the parabolic cylinder function of index $\nu$ \cite{Abramowitz}. The integration constants $A_{1,2}$ are such that
\begin{equation}
A_1 D_{\lambda/\beta}(Y) + A_2 D_{\lambda/\beta} (-Y) =0
\label{condinit}
\end{equation}
with $Y=\sqrt{2\beta/\Gamma}X$. After one performs the same integrations as in Sec. \ref{secjumpnoise}, Eq.\,(\ref{FokkerPlanckVP}) gives the two relations
\begin{equation}
\lambda_{X} f^{(0)}_{\lambda X}=\frac{\Gamma}{2} \left. \frac{\dd P^{\lambda}_{+}}{\dd x}\right|_{X}
\label{eqnthetaG1}
\end{equation}
\begin{equation}
\lambda_{X} f^{(1)}_{\lambda X}=\beta f^{(1)}_{\lambda X} + \frac{\Gamma}{2} X \left. \frac{\dd P^{\lambda}_{+}}{\dd x}\right|_{X}
\label{eqnthetaG2}
\end{equation}
Eliminating $\left.\frac{\dd P^{\lambda}_{+}}{\dd x}\right|_{X}$ from Eqs.\,(\ref{eqnthetaG1}) and (\ref{eqnthetaG2}) one finally gets
\begin{equation}
(f^{(1)}_{\lambda X}-X f^{(0)}_{\lambda X})\lambda_{X} = \beta f^{(1)}_{\lambda X}
\label{eqnthetaG}
\end{equation}
which is strictly equivalent to Eq.\,(\ref{eqntheta}). In the same way as in Sec. 2 it now follows that if $X=0$ then $\theta=\lambda^*_0=\beta$, which is the known result of Gaussian persistence \cite{MBCS,SMR,OCB}. 

Another remark can be made here. The solution of Eq\,(\ref{FokkerPlanckVP}) that vanishes at the origin is proportional to (with $A_1=-A_2=A$)
\begin{equation}
\tilde{P}(y)=A \ee ^{-y^2/4} (D _{\lambda/\beta} (y) - D _{\lambda/\beta} (-y) )
\end{equation}
which can be reduced to
\begin{equation}
\tilde{P}(y)=A' y \ee ^{-y^2/2}  \ _{1} F _{1} \left( (1-\lambda/\beta)/2,3/2,y^2/2 \right)
\end{equation}
where $A'$ is simply a numerical constant \cite{Abramowitz}. Using the result \cite{Gradshteyn}
\begin{equation}
\int_{0}^{\infty} \ee ^{-u} u^{1/2} \ _{1} F _{1} \left( (1-\nu)/2,3/2,u \right) \, \dd u =0 \qquad \nu > 1
\end{equation}
one has immediately $f^{(1)}_{\lambda 0}=0$ for any $\lambda > \beta$, which is a direct proof of Eq.\,(\ref{eqnonezero}) in the Gaussian case.


\subsection{Perturbation for small $X$}

Our method offers a faster way than the one presented in \cite{SMR}, in the Gaussian case, to calculate $\theta_X$ when $X$ is close to $\langle x \rangle$ (which is $0$ here).

As $P^{\beta}_+(x) = A' y \ee ^{-y^2/2}$ \cite{SMR}, one finds ${\langle x \rangle}_{\langle x \rangle}=\sqrt{\pi \Gamma/4\beta}$ and Eq.\,(\ref{thetapert}) gives immediately
\begin{equation}
\theta_X=\beta (1 + 2\sqrt{\frac{\beta}{\pi \Gamma}} X + \ldots) \, , \qquad X \to 0
\end{equation}
which is the well known perturbative result for the Gaussian persistence exponent \cite{SMR}.

\section{Application}

We give here some examples of Markovian but non-Gaussian processes defined by Eq.\,(\ref{langevin}).

\subsection{1D Percolation}
Let us consider an infinite 1D lattice. Each site can be occupied or empty. At an initial time, say $t=0$, we start with a totally empty lattice. Then, at each time step, one site is chosen randomly among non occupied ones and is then set occupied forever. The density of occupied sites $p$ is therefore an increasing function of time. It goes from $0$ to $1$ as time goes from $0$ to $\infty$ (in the limit of an infinite lattice, which is the case we consider here). We define clusters in a standard way, that is as sets of adjacent occupied sites. It is then possible to study percolation on this lattice as a dynamical process. de Freitas and dos Santos Lucena \cite{FreitasLucena} have already performed this study numerically.

The interest here is to consider a semi-infinite lattice and the cluster that contains its origin. When $p$ goes from 0 to 1, the size $s(p)$ of this cluster will grow from 0 to $\infty$. It is very easy to see that
\begin{equation}
\langle s \rangle = \frac{p}{1-p} \qquad \text{and}  \qquad \sqrt{\Delta s^2 } = \frac{\sqrt{p}}{1-p}
\end{equation}

The evolution equation for $s(p)$ is also quite simple. This random process is constant except when new sites are connected to the cluster which creates upward jumps of height $j$. Hence 
\begin{equation}
\frac{\dd s}{\dd p} = \sum_{k=0}^{\infty} j_k \delta (p-p_k)
\label{eqnmouvement}
\end{equation}
where $k$ is the jump index. If $\tilde{R}(j) \Delta p$ is the probability to have a jump $j$ when the density of occupied sites goes from $p$ to $p+\Delta p$ then 
\begin{equation}
\tilde{R}(j) \Delta p = p^{j-1} (1-p) \frac{\Delta p}{1-p}
\end{equation}
The quantity $\Delta p /(1-p)$ is simply the probability to make a jump during $\Delta p$. The normalised distribution law for the jumps is then
\begin{equation}
\tilde{R}(j) = p^{j-1} (1-p)
\end{equation}
and the density of jumps is $\rho(p)=1/(1-p)$. As $p$ tends to 1, the typical time scale of the dynamics of the process $s(p)$, which is $\sim 1/\rho$, goes to zero.
The aim is now to deal with a stationary process. If the new time scale is 
\begin{equation}
t=\log \frac{1}{1-p}
\end{equation}
the jump times $t_k$ are now uniformly distributed with a density 1 along the $t$ axis. Then we rescale the process $s(p)$, setting
\begin{equation}
x(t)=\frac{s(p)}{\sqrt{\Delta s^2}} \qquad \text{and} \qquad z_k=\frac{j_k}{\sqrt{\Delta s^2}}
\end{equation}
and we take the limit $p \to 1$ or equivalently $t \to \infty$. The jumps $z$ have now the probability density $R(z)$ so that $\tilde{R}(j)\Delta j =R(z)\Delta z$. In the limit $p \to 1$
\begin{equation}
R(z)=\ee ^{-z}
\end{equation}
If we write Eq.\,(\ref{eqnmouvement}) in terms of $x$, $t$, and $z$, the evolution equation is finally
\begin{equation}
\frac{\dd x}{\dd t} = - x(t) + \sum_{k=0}^{\infty} z_k \delta (t-t_k)
\label{eqnmouvementstat}
\end{equation}
in the limit $t \to \infty$. This gives a second example of noise of Eq.\,(\ref{noisedelta}) with an exponential distribution law for the jump height. The value of the persistence exponent is then $\theta=1$.


\subsection{Other examples}
A second example of noise defined by Eq\,(\ref{noisedelta}) is given by the study of random walks. The time $t(S)$ needed for the walker to visit $S$ distinct sites on a 1D lattice is a random process which, under the same kind of scaling as in the last subsection, satisfies Eq.\,(\ref{langevin}) \cite{GDH}. The jump height distribution $R(z)$ is then a power law with a cutoff. In this case the value of $\theta$ is $2$. We will not go further about this result in the present paper and the reader can see references \cite{GDH} and \cite{DH} for more details.

Finally the process $x(t)$ defined by Eq.\,(\ref{langevin}) is simply the electric signal that one would observe in an RC circuit receiving random charges due to, for instance, the presence of radioactive elements. The persistence exponent is then $\theta = 1 / RC$, that is the inverse relaxation time of the circuit.

\section{Conclusion and discussion}
\label{secconclusion}

With a very simple calculation we are able to find the persistence exponent of a large class of stationary Markov processes which includes the Gaussian case. As this quantity is universal within this entire class, it demonstrates the existence of universality classes related to non-Gaussian persistence exponents.

To go further one should study the properties of the master equation (\ref{Mastereqn}) because the level-dependent exponent $\theta_X$ is entirely determined by the knowledge of ${\langle x \rangle}_X$. However the nontrivial asymptotic behaviour $\theta_X \sim X \log X + X^{1/3} (\log X)^{2/3} + \ldots$, $X\rightarrow \infty$, found in Ref.\,\cite{DH} when the jumps are constant shows that this investigation is likely to be very difficult.


\section*{Acknowledgments}

The author thanks Professor H.~J. Hilhorst for numerous fruitful discussions  about the present work and also Professor L. dos Santos Lucena of the Universidade Federal do Rio Grande do Norte, Natal, Brazil for pointing out the relation between 1D percolation and noise with discrete upward jumps. 



\begin{thebibliography}{9}

\bibitem{majum}
S.~N. Majumdar,
\newblock {\it Current Science} {\bf 77} (1999) 370.

\bibitem{MBCS}
S.~N. Majumdar, A.~J. Bray, and S.~J. Cornell, C. Sire,
\newblock{\it Phys. Rev. Lett.} {\bf 77} (1996) 2867.

\bibitem{SMR}
C. Sire, S.~N. Majumdar, and A. R{\"u}dinger,
\newblock{\it Phys. Rev. E} {\bf 61} (2000) 1258.

\bibitem{OCB}
K. Oerding, S.~J. Cornell, and A.~J. Bray,
\newblock{\it Phys. Rev. E} {\bf 56} (1997) R25.

\bibitem{Abramowitz}
M. Abramowitz and I.~A. Stegun,
\newblock{\it Handbook of Mathematical Methods}
(Dover Publications, Inc., New York, 1965, \textit{ninth edition} 1972)

\bibitem{Gradshteyn}
I.~S. Gradshteyn and I.~M. Ryzhik,
\newblock{\it Table of integrals, series and products}
\newline {(Academic Press, 1965, \textit{fourth edition} 1980)}

\bibitem{GDH}
S.~R. Gomes J{\'u}nior, O. Deloubri{\`e}re, and H.J. Hilhorst,
\newblock unpublished.

\bibitem{DH}
O. Deloubri{\`e}re and H.J. Hilhorst,
\newblock {\it J. Phys. A: Math. Gen.} {\bf 33} (2000) 1993.

\bibitem{FreitasLucena}
J.~E. de Freitas and L. dos Santos Lucena,
\newblock{\it Physica A} {\bf 266} (1999) 81.


\end{thebibliography}
\end{document}